\documentclass[acmsmall,screen]{acmart}

\AtBeginDocument{%
  \providecommand\BibTeX{{%
    \normalfont B\kern-0.5em{\scshape i\kern-0.25em b}\kern-0.8em\TeX}}}

\copyrightyear{2023}
\acmYear{2023}
\setcopyright{acmlicensed}\acmConference[ETRA '23]{2023 Symposium on Eye Tracking Research and Applications}{May 30-June 2, 2023}{Tubingen, Germany}
\acmBooktitle{2023 Symposium on Eye Tracking Research and Applications (ETRA '23), May 30-June 2, 2023, Tubingen, Germany}
\acmPrice{15.00, Preprint Version}
\acmDOI{10.1145/3588015.3588404}
\acmISBN{979-8-4007-0150-4/23/05}

\usepackage{colortbl}
 \usepackage{hhline}
 
\begin{document}

\title[User-Interaction Based Reading Region Prediction]{Getting the Most from Eye-Tracking: User-Interaction Based Reading Region Estimation Dataset and Models}

\author{Ruoyan Kong}
\email{kong0135@umn.edu}
\affiliation{%
  \institution{University of Minnesota - Twin Cities}
  \country{USA}
}

\author{Ruixuan Sun}
\affiliation{%
  \institution{University of Minnesota - Twin Cities}
  \country{USA}
}
\author{Charles Chuankai Zhang}
\affiliation{%
  \institution{University of Minnesota - Twin Cities}
  \country{USA}
}
\author{Chen Chen}
\affiliation{%
  \institution{University of Minnesota - Twin Cities}
  \country{USA}
}
\author{Sneha Patri}
\affiliation{%
  \institution{University of Minnesota - Twin Cities}
  \country{USA}
}
\author{Gayathri Gajjela}
\affiliation{%
  \institution{University of Minnesota - Twin Cities}
  \country{USA}
}
\author{Joseph A. Konstan}
\email{konstan@umn.edu}
\affiliation{%
  \institution{University of Minnesota - Twin Cities}
  \country{USA}
}

\renewcommand{\shortauthors}{Kong, et al.}

\begin{abstract}
A single digital newsletter usually contains many messages (regions). Users' reading time spent on, and read level (skip/skim/read-in-detail) of each message is important for platforms to understand their users' interests, personalize their contents, and make recommendations. Based on accurate but expensive-to-collect eyetracker-recorded data, we built models that predict per-region reading time based on easy-to-collect Javascript browser tracking data. 

With eye-tracking, we collected 200k ground-truth datapoints on participants reading news on browsers. Then we trained machine learning and deep learning models to predict message-level reading time based on user interactions like mouse position, scrolling, and clicking. We reached 27\% percentage error in reading time estimation with a two-tower neural network based on user interactions only, against the eye-tracking ground truth data, while the heuristic baselines have around 46\% percentage error. We also discovered the benefits of replacing per-session models with per-timestamp models, and adding user pattern features. We concluded with suggestions on developing message-level reading estimation techniques based on available data.
\end{abstract}

\ccsdesc[500]{Human-centered computing~Collaborative and social computing systems and tools}
\ccsdesc[300]{Human-centered computing~Empirical studies in collaborative and social computing}

\keywords{ eye-tracking, reading region estimation, reading time, recommendation, personalization, neural network, bulk email}

\maketitle

\section{Introduction}
\begin{figure}[!htbp]
\centering
  \includegraphics[width=0.45\columnwidth]{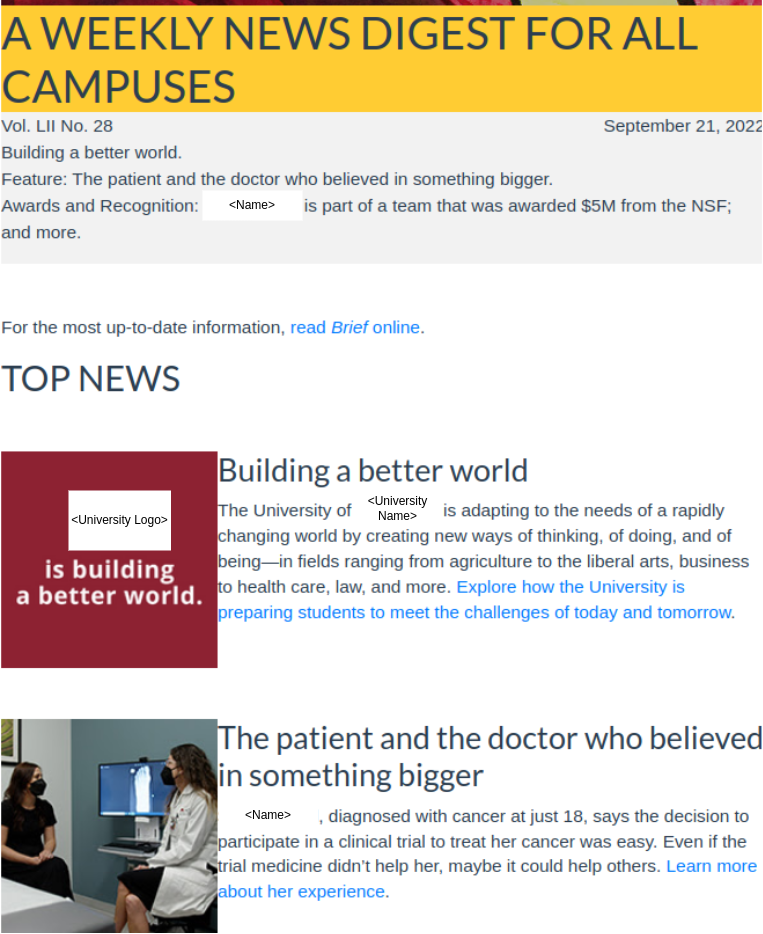}
  \caption{An example e-newsletter page of the study site with 30 messages. By message-level reading estimation, we estimate each message's (like ``Building a better world'')  reading time and read level for each user. }~\label{fig:sample_page}
\end{figure}

Digital newsletters (such as the one shown in Fig \ref{fig:sample_page}) include a variety of different \textbf{messages} within them \cite{kong2021learning, kong2022multi, economic_model}. We are interested in estimating whether \textbf{each user} read in detail, skimmed, or skipped each message (\textbf{read level}) to help senders understand their users' reading interests. Eye-tracking \cite{gibaldi2017evaluation, funke2016eye, othman2020eye} would provide a good solution, but users lack the hardware (and willingness) to provide that data. Instead, we rely on browser instrumentation to track time, scrolling, mouse clicks, etc. In this paper, we show how collecting a modest set of eye-tracking data and using it to train machine learning models for estimating reading behavior from browser logs can produce a fairly accurate classification of \textbf{message-level} reading behaviors.  

Our core research question is \textbf{how to estimate message-level reading time and read level based on each user's interactions with browsers}. We looked progressively at heuristics, logistic model and deep learning to identify the best estimation model, with following questions:

\noindent\textbf{Q1}: Do machine learning models, like the logistic model, perform better than the heuristic baselines?

\noindent\textbf{Q2:} Do neural network models perform better than the logistic model?

\noindent\textbf{Q3:} Do neural network models adjusted by user patterns perform better than the neural network without adjustment?

\noindent\textbf{Q4:} Are there performance differences between the neural network models that take the user / message features as input with those that take baseline predictions as input?

\noindent\textbf{Q5:} Are there performance differences between the models which make one estimation per timestamp with the models which make one estimation per reading session? \footnote{We defined a reading session as the duration between a user opens a newsletter and closes it or goes to other irrelevant tabs (which are not hyperlinks in the newsletter).}

We evaluated various models' performance on the users they had not seen in training by cross validation. We reached 27\% percentage error in reading time estimation with a two-tower neural network based on user interactions only, against the eye-tracking ground truth data, while the heuristic baselines have around 46\% percentage error. We found that neural networks' prediction error is decreased by 2\% when adding user pattern features; the models which make one estimation per second perform better than those which make one estimation per reading session. The proposed approach can enable platforms to make message-level reading estimations without extra devices and thus better personalize their contents. \footnote{The dataset and code are available at \url{https://github.com/ruoyankong/reading_region_prediction_dataset}.}


\section{Related Work}
\subsection{Reading Estimation Based on Eye-tracking Technology}
Eye-tracking technology refers to the applications, devices, and algorithms that use user's eye movement to catch their attention trace \cite{lai2013review}. It has been widely used in information processing tasks, including reading estimation of advertisement \cite{rayner2008eye}, literature \cite{augereau2016estimation}, social media \cite{namara2019you}, recommendations \cite{zhao2016gaze}, etc.

Remote eye-trackers and mobile eye-trackers are often used in reading behavior estimation. The remote eye-trackers use cameras, pupil center, and cornea reflection to track user's gaze position \cite{bohme2006remote}. Example eye-trackers in this category include Tobii-Pro-TX300, which contains separate eye-tracking units \footnote{https://www.tobii.com/products/discontinued/tobii-pro-tx300}, or GazeRecorder, which simply uses Webcam \footnote{https://gazerecorder.com/}. The mobile eye-trackers are often head-mounted devices that users need to wear and use a camera on the visual path to record the view, such as Tobii Pro Glasses \footnote{https://www.tobii.com/products/eye-trackers/wearables/tobii-pro-glasses-3}. 

Though eyetrackers provide relatively accurate data on reading estimation, they either ask users to wear/use specific devices or allow access to their webcams. These accesses are hard to get on a large scale for platforms, most of which only have access to browser interaction data. \textbf{Therefore there is a need to build message-level reading estimation models based on available browser data}.

\subsection{Reading Estimation Based on User Interactions}

Several studies have been done in estimating reading behaviors (e.g., time, focus area) based on user interactions. \citeauthor{seifert2017focus} studied which regions of the web pages users focus on when they read Wikipedia and found using single features (mouse position, paragraph position, mouse activity) cannot reach enough accuracy in estimating reading regions \cite{seifert2017focus}. \citeauthor{huang2012user} and \citeauthor{chen2001can} built linear models based on cursor position to estimate user's gaze coordinates during the web search / browsing \cite{huang2012user, chen2001can}.  \citeauthor{huang2012improving} and \citeauthor{10.1145/2766462.2767721} used cursor position in ranking search results \cite{huang2012improving, 10.1145/2766462.2767721} and found that 60\% of the tested queries were benefited by incorporating cursor data or significantly improve clicking prediction accuracy by 5\%.   \citeauthor{10.1145/2661829.2661909} recorded how users respond to online news and found the distance between mouse cursor and the corresponding news is significantly correlated with user's interestingness \cite{10.1145/2661829.2661909}.

However, most of these models did not consider user patterns. For example, the mouse position feature was treated the same for a user who moves mouse frequently versus a user who moves mouse infrequently \cite{lagun2015inferring}. \citeauthor{hauger2011using}'s work proposes to adjust reading estimations with user behavioral patterns --- but only a linear weight model was evaluated on the ground truth \cite{hauger2011using}. In fact, most of these studies focused on learning statistical correlations or only evaluated the proposed model's performance on groundtruth / heuristics. \textbf{We therefore found a need to evaluate various features and models' performance on reading estimation systematically}.

\section{Methods}
We defined \textbf{``reading estimation''} as 1) estimating reading time: how much time a user spends on reading each message per session (e.g., a user reads a message for 5 seconds); 2) classifying read level: whether a user skipped or skimmed a message, or read it in detail. Below, we introduced our data collection process (eye-tracking tests) and how we built features, models, and metrics.
\subsection{Eye-tracking Tests}
We conducted eye-tracking tests to collect interaction data and ground-truth labels. Participants were recruited through 5 mailing lists of the university employees / in-person contacts of the research team \footnote{This study is determined as non-human subject research by the university's IRB}. We excluded the university's communication professionals, as they might have processed the test messages. The selected participants were invited to the university's usability lab. 

Each participant then received 8 e-newsletters sampled from a diverse set of senders. The e-newsletters in the pool were selected randomly from the
university-wide e-newsletters sent in 2022, containing 3 to 30 messages. During the test, the participants' gaze positions were tracked by Tobii-TX300. They were asked to read as naturally as possible and read at least 1 message of each e-newsletter in detail. They can move mouse, scroll pages, and click on any hyperlinks. All these interactions were caught by browser Javascript and sent to the backend database (Google Firebase) in real-time. They can spend at most 30 minutes reading these e-newsletters and can leave anytime if they finish reading. The research team labeled the message located at the participants' gaze positions per second and calculated the reading time of each message as the ground truth. 


After each eye-tracking test, we retrained our models to check whether we got improvements in model performance. We stopped recruiting when we did not observe further improvements in the models' performance (see section 4). We conducted 9 eye-tracking tests finally, which resulted in a 200k dataset (one data point for each message's reading status per second).

\subsection{Features}
To consider which interaction features a model solely based on user interactions can use, we identified the data a browser can collect in a reading session. They can be classified into a 2x4 category: temporary/sessional features x pattern / user / message / baseline features.

\textbf{Temporary features} are collected per timestamp (each second for our case, e.g., the mouse position at time t). \textbf{Sessional features} are summaries of a reading session (e.g., the total number of seconds a message is visible on window in a reading session). \textbf{User features} represent user's status (e.g., mouse position). \textbf{Message features} represent message's status (e.g., message's position). \textbf{Pattern features} represent user's real-time behavioral patterns (e.g., mouse moving frequency until the latest timestamp). \textbf{Baseline features} represent the estimation given by the baselines --- the heuristics found to be correlated in the previous literature \cite{10.1145/1753846.1753976, seifert2017focus, huang2012user} (e.g., the share of a message on window). See the github repository for feature definitions.

\subsection{Models}
\begin{figure}[!htbp]
\centering
  \includegraphics[width=1\columnwidth]{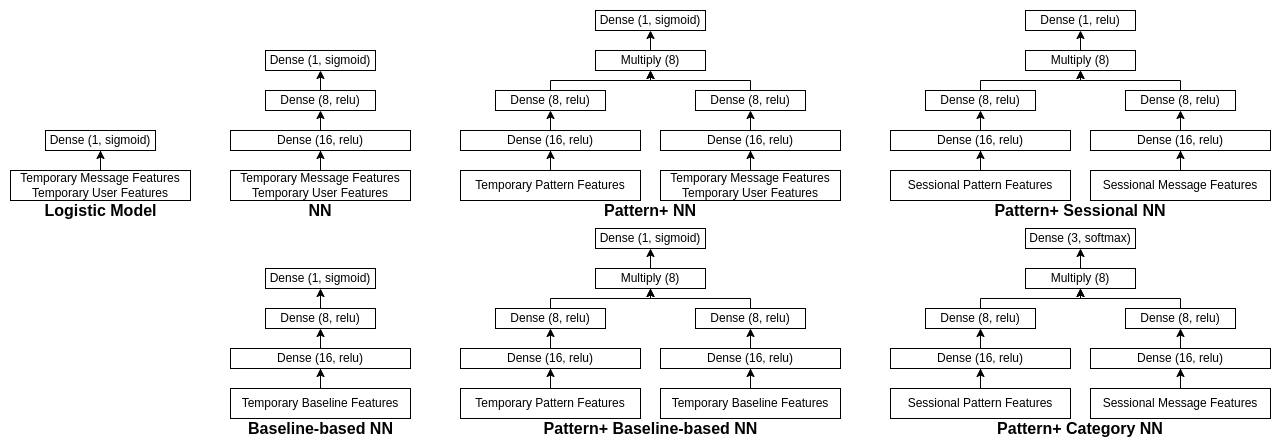}
  \caption{The estimation models. Logistic / NN / Pattern+ NN / Baselin-based NN / Pattern+ Baseline-based NN output $p_{m,t}$ per second to optimize binary loss. Pattern+ Sessional NN outputs $time_m$ per session to optimize absolute error loss. Pattern+ Category NN outputs $read\_level_m$ per session to optimize crossentropy loss.}~\label{fig:model}
\end{figure}

We considered a list of reading estimation models that use the interaction features above (Figure \ref{fig:model}).
\subsubsection{Per-Timestamp Models}
Per-timestamp models estimates the probability $p_{m,t}$ of a message $m$ being read at each timestamp $t$. We summarize per-timestamp models' estimations for each message at the end of each reading session to estimate the reading time and read level. Previous work found that message size, message position, and mouse position are usually found to be correlated with reading behavior \cite{10.1145/1753846.1753976, seifert2017focus, huang2012user}. Thus we started from 3 baselines: $p_{m,t}$ = message $m$'s current share of window area (\textbf{baseline 1}); $p_{m,t}$ is weighted by the reverse of $m$'s distance to the window center(\textbf{baseline2}); $p_{m,t} = 1$ if $m$ is closest to user mouse (\textbf{baseline 3}).

\noindent\textbf{Logistic Model}: it takes temporary message/user features as inputs, $p_{m,t}$ as outputs.

After that, we considered neural networks to learn more complex decision boundaries.

\noindent\textbf{Baseline-based NN}: A fully connected neural network which takes temporary baseline features as inputs and $p_{m,t}$ as outputs.

Then we tried using pattern features to learn which neurons' outputs are important.

\noindent\textbf{Pattern+ Baseline-based NN:} A two-tower NN that takes baseline features in one tower and pattern features in another, then merges these two towers (multiply), and finally outputs $p_{m,t}$.

We also tried inputting user temporary interaction features directly as below.

\noindent\textbf{Neural Network (NN)}: A fully connected NN which takes temporary message / user features.

\noindent\textbf{Pattern+ NN}: A two-tower NN that takes temporary message/user features in one tower and pattern features in another, then merges these two towers (multiply), and finally outputs $p_{m,t}$.

\noindent\textbf{Read Level:} The reading time $time_m$ of a message $m$ in a reading session $t_1$ to $t_2$ is defined as $time_m = \sum_{t_1\leq t \leq t_2}p_{m,t}$. Given message $m$'s number of words ($word_m$), its \textbf{$read\_level_m$ (skip/skim/read-in-detail)} for a user is separated by the reading speed 400 words / minute and 200 words / minute. These limits are selected based on the 70\% and 80\% comprehension levels in \cite{nocera2018rapid}. And \textbf{read} means a message is either skimmed or read-in-detail by a user.


\subsubsection{Per-Session Models}

Instead of making one estimation at each timestamp, we can also make one estimation per reading session.

\noindent\textbf{Pattern+ Sessional NN}:  A two-tower NN that takes sessional message features in one tower and sessional pattern features in another, then merges these two towers (multiply), and finally outputs reading time $time_m$ in a fully-connected ReLu layer.

\noindent\textbf{Pattern+ Category NN}: A two-tower NN that takes sessional message features in one tower and sessional pattern features in another, then merges these two towers (multiply), and finally outputs \textbf{$read\_level_m$} in a fully-connected softmax layer.

\subsubsection{Metrics} For per message in per reading session, we measured the error of estimating reading time and the accuracy of classifying read-level category (skip/skim/read-in-detail), see Table \ref{tab: metrics}.

\begin{table}[!htbp]
\centering
\caption{The Metrics' Definitions. Precision/recall is based on the specific skim/detail/read category.}~\label{tab: metrics}
\resizebox{0.85\textwidth}{!}{%
\begin{tabular}{|l|l|} 
\hline
\textbf{Metric}                                                                                                        & \textbf{Definition}                                                                                                                                                                                     \\ 
\hline
\begin{tabular}[c]{@{}l@{}}percentage error of \\reading time (per\_error)\end{tabular}                                & \begin{tabular}[c]{@{}l@{}}$per\_error = |true \ reading \ time - estimated \ reading \ time|/true \ reading \ time$\\when $true \ reading \ time \geq 10s$\end{tabular}                                \\ 
\hline
\begin{tabular}[c]{@{}l@{}}absolute error of \\reading time (abs\_error)\end{tabular}                                  & \begin{tabular}[c]{@{}l@{}}$abs\_error = |true \ reading \ time - estimated \ reading \ time|$\\when~$true \ reading \ time < 10s$\end{tabular}                                                      \\ 
\hline
\begin{tabular}[c]{@{}l@{}}accuracy of read-level \\(accuracy)\end{tabular}                                            & \begin{tabular}[c]{@{}l@{}}$accuracy = \sum_{m \in \{messages\}} I(estimated$ $category = true \ category) / |\{messages\}|$~\\$\{messages\}$ are all the messages from all the sessions.\end{tabular}  \\ 
\hline
\begin{tabular}[c]{@{}l@{}}precision of~read-level (skim\_precision,\\detail\_precision, read\_precision)\end{tabular} & e.g., $skim\_precision = \frac{\sum_{m \in \{messages\}} I(true \ category = skim) I(estimated \ category = skim)}{\sum_{m \in \{messages\}} I(estimated \ category = skim)}$                           \\ 
\hline
\begin{tabular}[c]{@{}l@{}}recall of read-level (skim\_recall, \\detail\_recall, read\_recall)\end{tabular}            & e.g., $skim\_recall = \frac{\sum_{m \in \{messages\}} I(true \ category = skim) I(estimated \ category = skim)} {\sum_{m \in \{messages\}} I(true \ category = skim)}$                                  \\
\hline
\end{tabular}
}
\end{table}

\section{Results}
After collecting and labeling the data, we trained the models and compared their performance. We used the Adam optimizer, a batch size of 64, 50 epochs with early stop. We sampled the split of train/validation/test set for 72 rounds --- for each round, one participant's data were used as test set, and the rest participant's data were split into train set and validation set with a ratio of 7:1 based on reading sessions. The weight of positive samples is set as 20, given that each newsletter contains approximately 20 messages on average, and the user is reading one message at any time. 

The average performance of all the proposed models is shown in Table \ref{tab:perf}, and the pair-wise t-tests' results between these models are shown in Table \ref{tab:compare}. Per-timestamp neural network models (Baseline-based NN, Pattern+ Baseline-based NN, NN, Pattern+ NN) have relatively higher overall performance compared to other models. Pattern+ Category NN's accuracy is only 47\%, which might indicate that making one estimation on the read level category per session does not utilize information efficiently. 

In short, \textbf{for the main research question, we can reach approximately 27\% percentage error in reading time estimation with models based on user interactions only}, compared to the ground truth of using an eyetracker, while the heuristic baselines reach around 43-46\% percentage error. Specifically, with Pattern+ NN, which uses user interaction features as input and adjusts their weights with user behavioral patterns, for reading time estimation, we reached 27\% in percentage error and 1.7 seconds in absolute error; for read level classification, we reached 79\% in read category precision and 68\% in read category recall.

\begin{table}[!htbp]
\centering
\caption{The average performance of all the proposed models on the testsets (the users in testsets are not in the corresponding train / validation sets). Read: the message is either skimmed or read-in-detail by the user.}~\label{tab:perf}
\scalebox{0.7}{
\begin{tabular}{|c|c|c|c|c|c|c|c|c|c|} 
\hline
\textbf{model}                                                        & \begin{tabular}[c]{@{}c@{}}\textbf{per\_}\\\textbf{error}\\\textbf{(\%)}\end{tabular} & \begin{tabular}[c]{@{}c@{}}\textbf{abs\_}\\\textbf{error}\\\textbf{(s)}\end{tabular} & \begin{tabular}[c]{@{}c@{}}\textbf{accu-}\\\textbf{racy}\\\textbf{(\%)}\end{tabular} & \begin{tabular}[c]{@{}c@{}}\textbf{skim\_}\\\textbf{precision}\\\textbf{(\%)}\end{tabular} & \begin{tabular}[c]{@{}c@{}}\textbf{skim\_}\\\textbf{recall}\\\textbf{(\%)}\end{tabular} & \begin{tabular}[c]{@{}c@{}}\textbf{detail\_}\\\textbf{precision}\\\textbf{(\%)}\end{tabular} & \begin{tabular}[c]{@{}c@{}}\textbf{detail\_}\\\textbf{recall}\\\textbf{(\%)}\end{tabular} & \begin{tabular}[c]{@{}c@{}}\textbf{read\_}\\\textbf{precision}\\\textbf{(\%)}\end{tabular} & \begin{tabular}[c]{@{}c@{}}\textbf{read\_}\\\textbf{recall}\\\textbf{(\%)}\end{tabular}  \\ 
\hline
Baseline1                                                             & 43\%                                                                                  & 2.1s                                                                                 & 83\%                                                                                 & 47\%                                                                                       & 32\%                                                                                    & 39\%                                                                                         & 27\%                                                                                      & 69\%                                                                                       & 50\%                                                                                     \\ 
\hline
Baseline2                                                             & 45\%                                                                                  & 2.5s                                                                                 & 81\%                                                                                 & 36\%                                                                                       & 24\%                                                                                    & 42\%                                                                                         & 31\%                                                                                      & 67\%                                                                                       & 48\%                                                                                     \\ 
\hline
Baseline3                                                             & 46\%                                                                                  & 2.3s                                                                                 & 83\%                                                                                 & 49\%                                                                                       & 43\%                                                                                    & 33\%                                                                                         & 40\%                                                                                      & 66\%                                                                                       & 59\%                                                                                     \\ 
\hline
\begin{tabular}[c]{@{}c@{}}Logistic \\Model\end{tabular}              & 38\%                                                                                  & 2.5s                                                                                 & 82\%                                                                                 & 45\%                                                                                       & 42\%                                                                                    & 64\%                                                                                         & 51\%                                                                                      & 64\%                                                                                       & 58\%                                                                                     \\ 
\hline
\begin{tabular}[c]{@{}c@{}}Baseline-\\based NN\end{tabular}           & 30\%                                                                                  & 1.8s                                                                                 & 86\%                                                                                 & 59\%                                                                                       & 51\%                                                                                    & 67\%                                                                                         & 65\%                                                                                      & 77\%                                                                                       & 68\%                                                                                     \\ 
\hline
\begin{tabular}[c]{@{}c@{}}Pattern+\\Baseline-\\based NN\end{tabular} & 28\%                                                                                  & 1.7s                                                                                 & 87\%                                                                                 & 63\%                                                                                       & 55\%                                                                                    & 69\%                                                                                         & 64\%                                                                                      & 79\%                                                                                       & 70\%                                                                                     \\ 
\hline
NN                                                                    & 27\%                                                                                  & 1.7s                                                                                 & 87\%                                                                                 & 65\%                                                                                       & 54\%                                                                                    & 71\%                                                                                         & 63\%                                                                                      & 81\%                                                                                       & 68\%                                                                                     \\ 
\hline
\begin{tabular}[c]{@{}c@{}}Pattern+ \\NN\end{tabular}                 & 27\%                                                                                  & 1.7s                                                                                 & 87\%                                                                                 & 64\%                                                                                       & 55\%                                                                                    & 71\%                                                                                         & 62\%                                                                                      & 79\%                                                                                       & 68\%                                                                                     \\ 
\hline
\begin{tabular}[c]{@{}c@{}}Pattern+\\Category NN\end{tabular}         & \textbackslash{}                                                                      & \textbackslash{}                                                                     & 47\%                                                                                 & 22\%                                                                                       & 69\%                                                                                    & 34\%                                                                                         & 3\%                                                                                       & 32\%                                                                                       & 72\%                                                                                     \\ 
\hline
\begin{tabular}[c]{@{}c@{}}Pattern+\\Sessional NN\end{tabular}        & 61\%                                                                                  & 3.1s                                                                                 & 78\%                                                                                 & 36\%                                                                                       & 26\%                                                                                    & 30\%                                                                                         & 10\%                                                                                      & 57\%                                                                                       & 34\%                                                                                     \\
\hline
\end{tabular}
}
\end{table}

We then compared the relative pairs of models' performance according to questions Q1 to Q5 (see section 1). The comparison results are shown in Table \ref{tab:compare}. We summarized our answers below.

\begin{table}[!htbp]
\centering
\caption{Model performance comparisons. Each cell is model1's metric value -\textgreater{} model2's metric value (pvalue). '*': pvalue $\leq$ 0.05; '.': pvalue $\leq$ 0.10. The pvalues were adjusted by the holm-sidak method \cite{cardillo2006holm}. }~\label{tab:compare}
\scalebox{0.7}{
\begin{tabular}{|l|l|l|c|c|c|c|c|c|c|c|c|} 
\hline
\begin{tabular}[c]{@{}l@{}}\textbf{que-}\\\textbf{stion}\end{tabular} & \textbf{model1}                                                 & \textbf{model2}                                                        & \begin{tabular}[c]{@{}c@{}}\textbf{per\_}\\\textbf{error}\\\textbf{(\%)}\end{tabular}  & \begin{tabular}[c]{@{}c@{}}\textbf{abs\_}\\\textbf{error}\\\textbf{(s)}\end{tabular}       & \begin{tabular}[c]{@{}c@{}}\textbf{accu-}\\\textbf{racy}\\\textbf{(\%)}\end{tabular}   & \begin{tabular}[c]{@{}c@{}}\textbf{skim\_}\\\textbf{precision}\\\textbf{(\%)}\end{tabular} & \begin{tabular}[c]{@{}c@{}}\textbf{skim\_}\\\textbf{recall}\\\textbf{(\%)}\end{tabular} & \begin{tabular}[c]{@{}c@{}}\textbf{detail\_}\\\textbf{precision}\\\textbf{(\%)}\end{tabular} & \begin{tabular}[c]{@{}c@{}}\textbf{detail\_}\\\textbf{recall}\\\textbf{(\%)}\end{tabular} & \begin{tabular}[c]{@{}c@{}}\textbf{read\_}\\\textbf{precision}\\\textbf{(\%)}\end{tabular} & \begin{tabular}[c]{@{}c@{}}\textbf{read\_}\\\textbf{recall}\\\textbf{(\%)}\end{tabular}  \\ 
\hhline{|============|}
Q1                                                   & Baseline1                                                       & \begin{tabular}[c]{@{}l@{}}Logistic \\Model\end{tabular}               & \begin{tabular}[c]{@{}c@{}}\textbf{43-\textgreater{}38}\\\textbf{(0.00*)}\end{tabular} & \begin{tabular}[c]{@{}c@{}}\textbf{2.1-\textgreater{}2.5}\\\textbf{(0.00*)}\end{tabular} & \begin{tabular}[c]{@{}c@{}}83-\textgreater{}82\\(0.50)\end{tabular}                    & \begin{tabular}[c]{@{}c@{}}47-\textgreater{}45\\(0.53)\end{tabular}                        & \begin{tabular}[c]{@{}c@{}}\textbf{32-\textgreater{}42}\\\textbf{(0.00*)}\end{tabular}  & \begin{tabular}[c]{@{}c@{}}\textbf{39-\textgreater{}64}\\\textbf{(0.00*)}\end{tabular}       & \begin{tabular}[c]{@{}c@{}}\textbf{27-\textgreater{}51}\\\textbf{(0.00*)}\end{tabular}    & \begin{tabular}[c]{@{}c@{}}69-\textgreater{}64\\(0.32)\end{tabular}                        & \begin{tabular}[c]{@{}c@{}}\textbf{50-\textgreater{}58}\\\textbf{(0.00*)}\end{tabular}   \\ 
\cline{2-12}
                                                                      & Baseline2                                                       & \begin{tabular}[c]{@{}l@{}}Logistic\\Model\end{tabular}                & \begin{tabular}[c]{@{}c@{}}\textbf{45-\textgreater{}38}\\\textbf{(0.00*)}\end{tabular} & \begin{tabular}[c]{@{}c@{}}2.5-\textgreater{}2.5\\(0.62)\end{tabular} & \begin{tabular}[c]{@{}c@{}}81-\textgreater{}82\\(0.28)\end{tabular} & \begin{tabular}[c]{@{}c@{}}\textbf{36-\textgreater{}45}\\\textbf{(0.06.)}\end{tabular}     & \begin{tabular}[c]{@{}c@{}}\textbf{24-\textgreater{}42}\\\textbf{(0.00*)}\end{tabular}  & \begin{tabular}[c]{@{}c@{}}\textbf{42-\textgreater{}64}\\\textbf{(0.00*)}\end{tabular}       & \begin{tabular}[c]{@{}c@{}}\textbf{31-\textgreater{}51}\\\textbf{(0.00*)}\end{tabular}    & \begin{tabular}[c]{@{}c@{}}67-\textgreater{}64\\(0.62)\end{tabular}                        & \begin{tabular}[c]{@{}c@{}}\textbf{48-\textgreater{}58}\\\textbf{(0.00*)}\end{tabular}   \\ 
\cline{2-12}
                                                                      & Baseline3                                                       & \begin{tabular}[c]{@{}l@{}}Logistic\\Model\end{tabular}                & \begin{tabular}[c]{@{}c@{}}\textbf{46-\textgreater{}38}\\\textbf{(0.00*)}\end{tabular} & \begin{tabular}[c]{@{}c@{}}2.3-\textgreater{}2.5\\(0.23)\end{tabular}                    & \begin{tabular}[c]{@{}c@{}}83-\textgreater{}82\\(0.60)\end{tabular}                    & \begin{tabular}[c]{@{}c@{}}49-\textgreater{}45\\(0.53)\end{tabular}                        & \begin{tabular}[c]{@{}c@{}}43-\textgreater{}42\\(0.95)\end{tabular}                     & \begin{tabular}[c]{@{}c@{}}\textbf{33-\textgreater{}64}\\\textbf{(0.00*)}\end{tabular}       & \begin{tabular}[c]{@{}c@{}}40-\textgreater{}51\\(0.40)\end{tabular}                       & \begin{tabular}[c]{@{}c@{}}66-\textgreater{}64\\(0.95)\end{tabular}                        & \begin{tabular}[c]{@{}c@{}}59-\textgreater{}58\\(0.95)\end{tabular}                      \\ 
                                                                      \cline{2-12}
                                                                      & Baseline1                                                       & NN                                                                     & \begin{tabular}[c]{@{}c@{}}\textbf{43-\textgreater{}27}\\\textbf{(0.00*)}\end{tabular} & \begin{tabular}[c]{@{}c@{}}\textbf{2.1-\textgreater{}1.7}\\\textbf{(0.00*)}\end{tabular} & \begin{tabular}[c]{@{}c@{}}\textbf{83-\textgreater{}87}\\\textbf{(0.00*)}\end{tabular} & \begin{tabular}[c]{@{}c@{}}\textbf{47-\textgreater{}65}\\\textbf{(0.00*)}\end{tabular}     & \begin{tabular}[c]{@{}c@{}}\textbf{32-\textgreater{}54}\\\textbf{(0.00*)}\end{tabular}  & \begin{tabular}[c]{@{}c@{}}\textbf{39-\textgreater{}71}\\\textbf{(0.00*)}\end{tabular}       & \begin{tabular}[c]{@{}c@{}}\textbf{27-\textgreater{}63}\\\textbf{(0.00*)}\end{tabular}    & \begin{tabular}[c]{@{}c@{}}\textbf{69-\textgreater{}81}\\\textbf{(0.00*)}\end{tabular}     & \begin{tabular}[c]{@{}c@{}}\textbf{50-\textgreater{}68}\\\textbf{(0.00*)}\end{tabular}   \\ 

\hline
Q2                                                                    & \begin{tabular}[c]{@{}l@{}}Logistic \\Model\end{tabular}        & NN                                                                     & \begin{tabular}[c]{@{}c@{}}\textbf{38-\textgreater{}27}\\\textbf{(0.00*)}\end{tabular} & \begin{tabular}[c]{@{}c@{}}\textbf{2.5-\textgreater{}1.7}\\\textbf{(0.00*)}\end{tabular} & \begin{tabular}[c]{@{}c@{}}\textbf{82-\textgreater{}87}\\\textbf{(0.00*)}\end{tabular} & \begin{tabular}[c]{@{}c@{}}\textbf{45-\textgreater{}65}\\\textbf{(0.00*)}\end{tabular}     & \begin{tabular}[c]{@{}c@{}}\textbf{42-\textgreater{}54}\\\textbf{(0.00*)}\end{tabular}  & \begin{tabular}[c]{@{}c@{}}\textbf{64-\textgreater{}71}\\\textbf{(0.01*)}\end{tabular}                          & \begin{tabular}[c]{@{}c@{}}\textbf{51-\textgreater{}63}\\\textbf{(0.00*)}\end{tabular}    & \begin{tabular}[c]{@{}c@{}}\textbf{64-\textgreater{}81}\\\textbf{(0.00*)}\end{tabular}     & \begin{tabular}[c]{@{}c@{}}\textbf{58-\textgreater{}68}\\\textbf{(0.00*)}\end{tabular}   \\ 
\hline
Q3                                                & NN                                                              & \begin{tabular}[c]{@{}l@{}}Pattern+ \\NN\end{tabular}                  & \begin{tabular}[c]{@{}c@{}}27-\textgreater{}27\\(0.99)\end{tabular}                    & \begin{tabular}[c]{@{}c@{}}1.7-\textgreater{}1.7\\(0.76)\end{tabular}                    & \begin{tabular}[c]{@{}c@{}}87-\textgreater{}87\\(0.97)\end{tabular}                    & \begin{tabular}[c]{@{}c@{}}65-\textgreater{}64\\(0.98)\end{tabular}                        & \begin{tabular}[c]{@{}c@{}}54-\textgreater{}55\\(0.97)\end{tabular}                     & \begin{tabular}[c]{@{}c@{}}71-\textgreater{}71\\(0.99)\end{tabular}                          & \begin{tabular}[c]{@{}c@{}}63-\textgreater{}62\\(0.99)\end{tabular}                       & \begin{tabular}[c]{@{}c@{}}81-\textgreater{}79\\(0.48)\end{tabular}                        & \begin{tabular}[c]{@{}c@{}}68-\textgreater{}68\\(0.99)\end{tabular}                      \\ 
\cline{2-12}
                                                                      & \begin{tabular}[c]{@{}l@{}}Baseline-\\based NN\end{tabular}     & \begin{tabular}[c]{@{}l@{}}Pattern+ \\Baseline-\\based NN\end{tabular} & \begin{tabular}[c]{@{}c@{}}\textbf{30-\textgreater{}28}\\\textbf{(0.00*)}\end{tabular}                    & \begin{tabular}[c]{@{}c@{}}\textbf{1.8-\textgreater{}1.7}\\\textbf{(0.01*)}\end{tabular} & \begin{tabular}[c]{@{}c@{}}86-\textgreater{}87\\(0.17)\end{tabular}                    & \begin{tabular}[c]{@{}c@{}}59-\textgreater{}63\\(0.12)\end{tabular}     & \begin{tabular}[c]{@{}c@{}}\textbf{51-\textgreater{}55}\\\textbf{(0.10.)}\end{tabular}  & \begin{tabular}[c]{@{}c@{}}67-\textgreater{}69\\(0.24)\end{tabular}                          & \begin{tabular}[c]{@{}c@{}}65-\textgreater{}64\\(0.68)\end{tabular}                       & \begin{tabular}[c]{@{}c@{}}77-\textgreater{}79\\(0.24)\end{tabular}                        & \begin{tabular}[c]{@{}c@{}}68-\textgreater{}70\\(0.24)\end{tabular}   \\ 
\hline
Q4                                                                    & \begin{tabular}[c]{@{}l@{}}Pattern+ \\NN\end{tabular}           & \begin{tabular}[c]{@{}l@{}}Pattern+ \\Baseline-\\based NN\end{tabular} & \begin{tabular}[c]{@{}c@{}}27-\textgreater{}28\\(0.86)\end{tabular}                    & \begin{tabular}[c]{@{}c@{}}1.7-\textgreater{}1.7\\(0.92)\end{tabular}                    & \begin{tabular}[c]{@{}c@{}}87-\textgreater{}87\\(0.73)\end{tabular}                    & \begin{tabular}[c]{@{}c@{}}64-\textgreater{}63\\(0.86)\end{tabular}                        & \begin{tabular}[c]{@{}c@{}}55-\textgreater{}55\\(0.92)\end{tabular}                     & \begin{tabular}[c]{@{}c@{}}71-\textgreater{}69\\(0.86)\end{tabular}                          & \begin{tabular}[c]{@{}c@{}}62-\textgreater{}64\\(0.92)\end{tabular}                       & \begin{tabular}[c]{@{}c@{}}79-\textgreater{}79\\(0.92)\end{tabular}                        & \begin{tabular}[c]{@{}c@{}}68-\textgreater{}70\\(0.66)\end{tabular}   \\ 
\hline
Q5                                                  & \begin{tabular}[c]{@{}l@{}}Pattern+ \\Sessional NN\end{tabular} & \begin{tabular}[c]{@{}l@{}}Pattern+ \\NN\end{tabular}                  & \begin{tabular}[c]{@{}c@{}}\textbf{61-\textgreater{}27}\\\textbf{(0.00*)}\end{tabular} & \begin{tabular}[c]{@{}c@{}}\textbf{3.1-\textgreater{}1.7}\\\textbf{(0.00*)}\end{tabular} & \begin{tabular}[c]{@{}c@{}}\textbf{78-\textgreater{}87}\\\textbf{(0.00*)}\end{tabular} & \begin{tabular}[c]{@{}c@{}}\textbf{36-\textgreater{}65}\\\textbf{(0.00*)}\end{tabular}     & \begin{tabular}[c]{@{}c@{}}\textbf{26-\textgreater{}55}\\\textbf{(0.00*)}\end{tabular}  & \begin{tabular}[c]{@{}c@{}}\textbf{30-\textgreater{}71}\\\textbf{(0.00*)}\end{tabular}       & \begin{tabular}[c]{@{}c@{}}\textbf{10-\textgreater{}62}\\\textbf{(0.00*)}\end{tabular}    & \begin{tabular}[c]{@{}c@{}}\textbf{57-\textgreater{}79}\\\textbf{(0.00*)}\end{tabular}     & \begin{tabular}[c]{@{}c@{}}\textbf{34-\textgreater{}68}\\\textbf{(0.00*)}\end{tabular}   \\ 
\cline{2-12}
                                                                      & \begin{tabular}[c]{@{}l@{}}Pattern+ \\Category NN\end{tabular}  & \begin{tabular}[c]{@{}l@{}}Pattern+ \\NN\end{tabular}                  & \textbackslash{}                                                                       & \textbackslash{}                                                                           & \begin{tabular}[c]{@{}c@{}}\textbf{47-\textgreater{}87}\\\textbf{(0.00*)}\end{tabular} & \begin{tabular}[c]{@{}c@{}}\textbf{22-\textgreater{}65}\\\textbf{(0.00*)}\end{tabular}     & \begin{tabular}[c]{@{}c@{}}\textbf{69-\textgreater{}55}\\\textbf{(0.01*)}\end{tabular}  & \begin{tabular}[c]{@{}c@{}}\textbf{34-\textgreater{}71}\\\textbf{(0.01*)}\end{tabular}       & \begin{tabular}[c]{@{}c@{}}\textbf{3-\textgreater{}62}\\\textbf{(0.00*)}\end{tabular}     & \begin{tabular}[c]{@{}c@{}}\textbf{32-\textgreater{}79}\\\textbf{(0.00*)}\end{tabular}     & \begin{tabular}[c]{@{}c@{}}72-\textgreater{}68\\(0.45)\end{tabular}   \\
\hline
\end{tabular}
}
\end{table}

\noindent\textbf{Q1:} Do machine learning models perform better than the heuristic baselines?

Yes, the logistic model and neural network models perform better than the heuristic baselines on most of the numerical and classification metrics. For example, compared to baseline 1 (heuristic estimation based on window share), NN significantly improved per\_error from 43\% to 27\%, read\_precision from 69\% to 81\%, read\_recall from 50\% to 68\%, etc. Though the logistic model does not perform better than baseline1 on abs\_error, the NN model performs better than the baselines on all the metrics.

\noindent\textbf{Q2:} Do neural network models perform better than the logistic model?

Yes, the neural network model performs better than the logistic model on almost all the numerical and classification metrics (similar performance on detail\_precision). For example, compared to the logistic model, NN significantly improves per\_error from 38\% to 27\%, abs\_error from 2.5s to 1.7s, accuracy from 82\% to 87\%, etc.

\noindent\textbf{Q3:} Do neural network models adjusted by user patterns 
perform better than the single-tower neural network models?

Yes for the NN that takes baselines as input but not for the NN that takes message/user features as input. The baseline neural network adjusted by user pattern features further improves the baseline neural network on the absolute error and skim prediction (marginally) significantly: per\_error (30\% to 28\%), skim\_recall (51 to 55\%). There is no significant performance difference between NN and Pattern+ NN.

\noindent\textbf{Q4:} Are there performance differences between the neural network models that take the user /
message features as input with those that take baseline features as input?

No, the pattern+ neural network that takes baseline features as input performs similarly to the pattern+ NN that takes user / message features as input.

\noindent\textbf{Q5:} Are there performance differences between the models which make one estimation per timestamp with the models which make one estimation per reading session?

Yes, the models that make one estimation for each reading session perform significantly worse than the models that make one estimation each second (see Table \ref{tab:compare}). For example, Pattern+ Sessional NN's per\_error is 61\% compared to Pattern+ NN's 27\%, accuracy is 78\% versus Pattern+ NN's 87\%.

Here we reported the trends of our models' performance during the data collection process. We recruited participants in 4 rounds. After each round, we retrained the models and evaluated their performance with the training/testing process above. Table \ref{tab:trend} is the performance trend of the Pattern+ Baseline-based NN model. In these 4 rounds, we see improvements mainly in model per\_error (36\% to 28\%), abs\_error (2.1s to 1.7s). We stopped when we did not observe further improvements on per\_error and abs\_error.
\begin{table}[!htbp]
\centering
\caption{The performance trend of the Pattern+ Baseline-based NN model (its average performance in the train/test splits after each round of the data collection process).}~\label{tab:trend}
\scalebox{0.7}{
\begin{tabular}{|c|c|c|c|c|c|c|c|c|c|c|} 
\hline
\multicolumn{1}{|l|}{\textbf{round}} & \multicolumn{1}{l|}{\begin{tabular}[c]{@{}l@{}}\textbf{\#parti-}\\\textbf{cipants}\end{tabular}} & \begin{tabular}[c]{@{}c@{}}\textbf{per\_}\\\textbf{error}\\\textbf{(\%)}\end{tabular} & \begin{tabular}[c]{@{}c@{}}\textbf{abs\_}\\\textbf{error}\\\textbf{(s)}\end{tabular} & \begin{tabular}[c]{@{}c@{}}\textbf{accu-}\\\textbf{racy}\\\textbf{(\%)}\end{tabular} & \begin{tabular}[c]{@{}c@{}}\textbf{skim\_}\\\textbf{precision}\\\textbf{(\%)}\end{tabular} & \begin{tabular}[c]{@{}c@{}}\textbf{skim\_}\\\textbf{recall}\\\textbf{(\%)}\end{tabular} & \begin{tabular}[c]{@{}c@{}}\textbf{detail\_}\\\textbf{precision}\\\textbf{(\%)}\end{tabular} & \begin{tabular}[c]{@{}c@{}}\textbf{detail\_}\\\textbf{recall}\\\textbf{(\%)}\end{tabular} & \begin{tabular}[c]{@{}c@{}}\textbf{read\_}\\\textbf{precision}\\\textbf{(\%)}\end{tabular} & \begin{tabular}[c]{@{}c@{}}\textbf{read\_}\\\textbf{recall}\\\textbf{(\%)}\end{tabular}  \\ 
\hhline{|===========|}
1                                    & 2                                                                                                & 36\%                                                                                  & 2.1s                                                                                 & 87\%                                                                                 & 55\%                                                                                       & 54\%                                                                                    & 65\%                                                                                         & 65\%                                                                                      & 70\%                                                                                       & 67\%                                                                                     \\ 
\hline
2                                    & 5                                                                                                & 30\%                                                                                  & 2.0s                                                                                 & 84\%                                                                                 & 61\%                                                                                       & 52\%                                                                                    & 39\%                                                                                         & 38\%                                                                                      & 74\%                                                                                       & 63\%                                                                                     \\ 
\hline
3                                    & 7                                                                                                & 28\%                                                                                  & 1.7s                                                                                 & 86\%                                                                                 & 67\%                                                                                       & 52\%                                                                                    & 65\%                                                                                         & 67\%                                                                                      & 80\%                                                                                       & 66\%                                                                                     \\ 
\hline
4                                    & 9                                                                                                & 28\%                                                                                  & 1.7s                                                                                 & 87\%                                                                                 & 63\%                                                                                       & 55\%                                                                                    & 69\%                                                                                         & 64\%                                                                                      & 79\%                                                                                       & 70\%                                                                                     \\
\hline
\end{tabular}
}
\end{table}

\section{Discussion}

\subsection{Per-Timestamp and Per-Session Models.}
We found that the models which make one estimation per second (and summarize the estimations at the end of each session) perform significantly better than the models which make one estimation per session. The reason that per-timestamp models outperform per-session models might be that the per-session models fail to utilize a large part of the information that can be collected per timestamp. In the previous work on learning user's reading interests (either through duration data collected by eye-trackers or interaction data), the estimations are often made per session instead of per timestamp \cite{10.1145/2661829.2661909,augereau2016estimation,10.1145/2766462.2767721, li2017towards}. \textbf{Therefore we suggested future work on reading estimation tasks to evaluate both the per-timestamp and per-session models.} Meanwhile, as per-timestamp models need more computing resources compared to per-session models, future studies can also consider online learning \cite{lee2019intermittent, kong2021nimblelearn} to balance model performance and complexity.

\subsection{Improve accuracy by more features and more data.}We also found that users' behavioral pattern features helped the Baseline-based neural network to further decrease its error on reading time estimation. This result shows that user contexts could be used in improving reading estimation. Besides user behavioral patterns, other potential contexts to be considered included user's intent \cite{kong2021virtual} (e.g., search queries \cite{he2023que2engage, he2023hiercat}), preference (e.g., user's preference for movie genres \cite{zhao2016gaze, zhao2016group}), item's features (e.g., product ratings \cite{shahriar2020online, jin2021estimating}, movies tags \cite{aridor2022economics}).

\subsection{Limitations and Future Work}
We only collected eye-tracking data on 9 users therefore the dataset we collected might not catch enough variance on user patterns. Future work should look at where the value of additional users starts to decay significantly.  


\section{Conclusion}
We studied how to estimate message-level reading time and
read level of digital newsletters using user interactions extracted by browser JavaScript (e.g., mouse scroll, click, and hover actions). We applied the accurate but hard-to-scale eye-tracking data to learn the association between user interactions and reading time. We conducted 9 eye-tracking tests and collected 200k per-second participants' reading (gaze position) and interaction datapoints. The dataset will be public. 

We reached a 27\% percentage error in reading time estimation with a two-tower neural network based on user interactions only, while the heuristic baselines have around 43\% to 46\% percentage error. We found that 1) the per-timestamp models perform significantly better than the per-session models; 2) the neural networks' performance might be further improved by adjusting them with user pattern features. This paper 1) provides examples of generalizing eye-tracking data to build reading estimation approaches that use browser-extracted features only and thus can be applied on a large scale; 2) give insights for future reading estimation studies on designing features / models.

\begin{acks}
This work was supported by the National Science Foundation under grant CNS-2016397. We thank the University of Minnesota's Usability Lab's User Experience Analysts, Nick Rosencrans, and David Rosen, for their assistance with eye-tracking technology.
\end{acks}

\bibliographystyle{ACM-Reference-Format}
\bibliography{sample-base}


\end{document}